\documentstyle[12pt,epsfig]{article}
\newcommand{\beq}{\begin{equation}}
\newcommand{\eeq}{\end{equation}}
\newcommand{\beqn}{\begin{eqnarray}}
\newcommand{\eeqn}{\end{eqnarray}}

\newcommand{\stackm}{\stackrel{\scriptstyle <}{{ }_{\sim}}}  
\begin{document}

\thispagestyle{empty}
\def\pubnum{433}
\def\data{February, 1998}
\begin{flushright}
{\parbox{3.5cm}{
UAB-FT-437

February, 1998

hep-ph/9802363
}}
\end{flushright}
\vspace{3cm}
\begin{center}
\begin{large}
\begin{bf}
$t\rightarrow W^+\,b$ AND $t\rightarrow H^+\,b$
AT THE QUANTUM LEVEL\\ IN THE MSSM\\
\end{bf}
\end{large}
\vspace{1cm}
Joan SOL\`A\footnote{Talk presented at the {\it International
Workshop on Quantum Effects in the MSSM}\,, Barcelona,  
September 9-13, 1997. To appear in the Proceedings.}\\

\vspace{0.25cm} 
Grup de F\'{\i}sica Te\`orica\\ 
and\\ 
Institut de F\'\i sica d'Altes Energies\\ 
\vspace{0.25cm} 
Universitat Aut\`onoma de Barcelona\\
08193 Bellaterra (Barcelona), Catalonia, Spain\\
\end{center}
\vspace{0.3cm}
\hyphenation{super-symme-tric re-nor-ma-li-za-tion}
\hyphenation{com-pe-ti-ti-ve}
\begin{center}
{\bf ABSTRACT}
\end{center}
\begin{quotation}
\noindent
We compare the standard top quark decay and the
charged Higgs decay of the top quark,
$t\rightarrow W^+\,b$ and $t\rightarrow H^+\,b$,
at the quantum level 
in the MSSM. While the SUSY loop corrections to the standard
top quark decay are only of a few percent,
it turns out that $t\rightarrow H^+\,b$ is a most promising
candidate for carrying large quantum SUSY signatures. 
As a result, the $(\tan\beta, M_{H^{\pm}})$ exclusion plots 
presented by the CDF Collaboration should be
thoroughly revised in the
light of the MSSM.
\end{quotation}
  
\newpage

\baselineskip=6.5mm  %(FOR PREPRINT)
Theoretically, Supersymmetry (SUSY) is perhaps the only
known framework  beyond the 
Standard Model (SM) which is capable of extending
non-trivially the quantum field
theoretical structure of the conventional
strong and electroweak interactions while keeping all the
necessary ingredients insuring internal consistency, such as
gauge invariance and renormalizability.
In particular, the Minimal Supersymmetric
Standard Model (MSSM)\,\cite{Gunion,WdeBoer} has been able
to accommodate all known high precision
measurements to a similar degree
of significance as the Standard Model \,\cite{WdeBoer}.
Remarkably, SUSY has been able to survive over the years and
it has become a ``fact'' of live for many physicists.
Most likely this situation  will remain invariable 
at least until the Tevatron II and LHC eras have explored
in full the experimental feasibility of the MSSM.

In this talk I propose to dwell on the supersymmetric
phenomenology of
top quark decays with an eye on the Tevatron and LHC
phenomenological capabilities.
Among the relevant MSSM top quark decays potentially carrying
a direct or indirect SUSY signature, the following two-body 
modes stand out:
\beqn
{\rm i)} &&t\rightarrow W^+\,b\,\nonumber\\
{\rm ii)} &&t\rightarrow H^+\,b\,\nonumber\\
{\rm iii}) &&t\rightarrow \tilde{t}_a\,\chi^0_{\alpha},\nonumber\\
{\rm iv}) &&t\rightarrow \tilde{b}_a\,\chi^+_i,\nonumber\\
{\rm v}) &&t\rightarrow \tilde{t}_a\,\tilde{g}\,.
\label{eq:decays}
\eeqn
Therein, $\tilde{t}_{a}$, $\tilde{b}_{a}$, $\chi^+_i$,
$\chi^0_{\alpha}$, $\tilde{g}_r$ ($a,i=1,2;\,\alpha=1,2,...,4$;\,
$r=1,2,...,8$)
denote stop, sbottom, chargino, neutralino and gluino ``sparticles'',
respectively. 
(Also quite a few three-body decays are possible and have been 
studied\,\cite{Guasch1}.)
Of course, decay i) is the SM top quark decay, and decay ii) into 
a charged Higgs need not to be a SUSY decay. However, by studying
the possible MSSM quantum effects on these decay modes one may hope
to unveil indirect traces of the underlying SUSY dynamics. 
On the other hand the last three decays in (\ref{eq:decays})
do carry an explicit SUSY signature.
In general also these decays may require
a higher order treatment, the reason being that some of
the final state signatures, after the sparticles have decayed into 
conventional particles and the LSP (typically the lightest neutralino
$\chi_1^0$), they may well mimic the standard top
quark decay.
For example, decay iii) may lead to a signature similar to
the standard top quark decay into the final states 
$b\,l^+\,\nu$ or $b+2\,{\rm jets}$; for the stop could
decay into $\chi^+_i\,b$, and subsequently yield the chain
$\chi^+_i\rightarrow\chi^0_1\,W^*
\rightarrow\chi^0_1\,l^+\,\nu$ or $\chi^0_1+2\,{\rm jets}$. 
Therefore, a detailed treatment of these direct SUSY modes
is in principle desirable to help disentangling the nature of the 
complicated final configurations and to enable a reliable determination
of the top quark cross-section within the MSSM.
Barring a light gluino window, which is nowadays harder and harder
to maintain, current limits on squark and gluino masses already rule out
decay v) and most likely also decay iv). However, even keeping alive
this last decay, unfortunately the typical size of the corrections to
the two processes iii) and iv) is not too significant  (at the ten
per cent level at most\,\cite{Djouadi}). While this would amply suffice in
a high precision machine such as LEP, nevertheless for measurements to be 
performed in a hadron environment it is probably not enough to be detected.

The fact that sparticles seem to be rather heavy makes 
direct SUSY searches more and more difficult. For this reason
it may be advisable to hunt for
``quantum signatures'' by means of the 
indirect method of high precision measurements of
less exotic, and so more manageable, processes. 
Thus we shall report here on the behaviour of the more conventional
decays i) and ii) at the quantum level.
To start with, we recall that the supersymmetric
strong (SUSY-QCD) and the supersymmetric electroweak (SUSY-EW) 
corrections to the standard top quark decay 
$t\rightarrow W^+\,b$
are well understood\,\cite{GJSH}. The leading one-loop vertex functions
are shown in Fig.\,1 (taking the external dashed line as the $W^+$).  
The possibility of large non-standard quantum effects lies to a great extent
on the influence of the parameter $\tan\beta$\,\cite{Hunter}. 
In supersymmetric theories like the MSSM, $\tan\beta$ 
enters the top and bottom quark Yukawa couplings
of the superpotential through $1/\sin\beta$ and $1/\cos\beta$,
respectively:
\beq
h_t={g\,m_t\over \sqrt{2}\,M_W\,\sin{\beta}}\,,\
h_b={g\,m_b\over \sqrt{2}\,M_W\,\cos{\beta}}\,,
\eeq
and therefore one may expect an enhancement of the
Yukawa couplings as compared to the gauge couplings both at low
and high $\tan\beta$. 
Notice that the bottom-quark Yukawa coupling may counterbalance the
smallness of the bottom mass at the expense of a large value of
$\tan\beta$. 

Apart from $\tan\beta$, the basic free parameters of our
analysis concerning the electroweak sector are contained in the
stop and sbottom mass matrices ($\tilde{q}=\tilde{t},\tilde{b}$):
\begin{equation}
{\cal M}_{\tilde{q}}^2 =\left(\begin{array}{cc}
 {\cal M}_{11}^2 & {\cal M}_{12}^2 
\\ {\cal M}_{12}^2 &{\cal M}_{22}^2 \,.
\end{array} \right)\,,
\label{eq:stopmatrix}
\end{equation}
with 
\begin{eqnarray}
{\cal M}_{11}^2 &=&M_{\tilde{q}_L}^2+m_q^2\nonumber\\
&+&\cos{2\beta}(T^3_q-Q_q\,\sin^2\theta_W)\,M_Z^2\,,\nonumber\\
{\cal M}_{22}^2 &=&M_{\tilde{q}_R}^2+m_q^2\nonumber\\
&+&Q_q\,\cos{2\beta}\,\sin^2\theta_W\,M_Z^2\,,\nonumber\\
{\cal M}_{12}^2 &=&m_q\, M_{LR}^q\,,\nonumber\\
M_{LR}^{\{t,b\}}&=&A_{\{t,b\}}-\mu\{\cot\beta,\tan\beta\}\,. \ \ \ \
\end{eqnarray}
We denote by $m_{\tilde{t}_1}$
and $m_{\tilde{b}_1}$ the lightest stop and sbottom
masses.

%%%%%%%%%%%%%%%%%%%%%%%%%%%%%%%%%%%%%%%%%%%%%%%%%%%%%%%%%%%%%%%%%%%%%%%
%Fig.1
\begin{figure}
\centering
\mbox{\epsfig{file=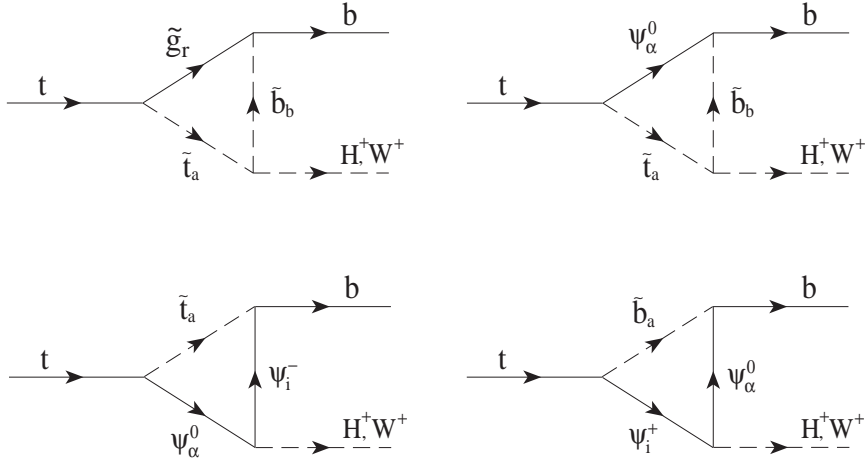,width=11.5cm}}
%%%
%%%%%%  Figures Centrades Verticalment
%%%
%\noindent
\caption
{SUSY-QCD and SUSY-EW one-loop vertices for 
$t\rightarrow W^+\,b$ and $t\rightarrow H^+\,b$.
The EW ``inos'' $\Psi_{i,\alpha}$ are unphysical
mass-eigenstates related
to the physical states $\chi_{i,\alpha}$ (Cf. Ref.[3]).}
%\protect{\cite{Guasch1}}.}
%%\end{center}
\end{figure}
%%%%%%%%%%%%%%%%%%%%%%%%%%%%%%%%%%%%%%%%%%%%%%%%%%%%%%%%%%%%%%%%%%%%%%%%%%%% 

%%%%%%%%%%%%%%%%%%%%%%%%%%%%%%%%%%%%%%%%%%%%%%%%%%%%%%%%%%%%%%%%%%%%%%%
%Fig.2
\begin{figure}
\centering
        \begin{tabular}{cc}
          \mbox{\epsfig{file=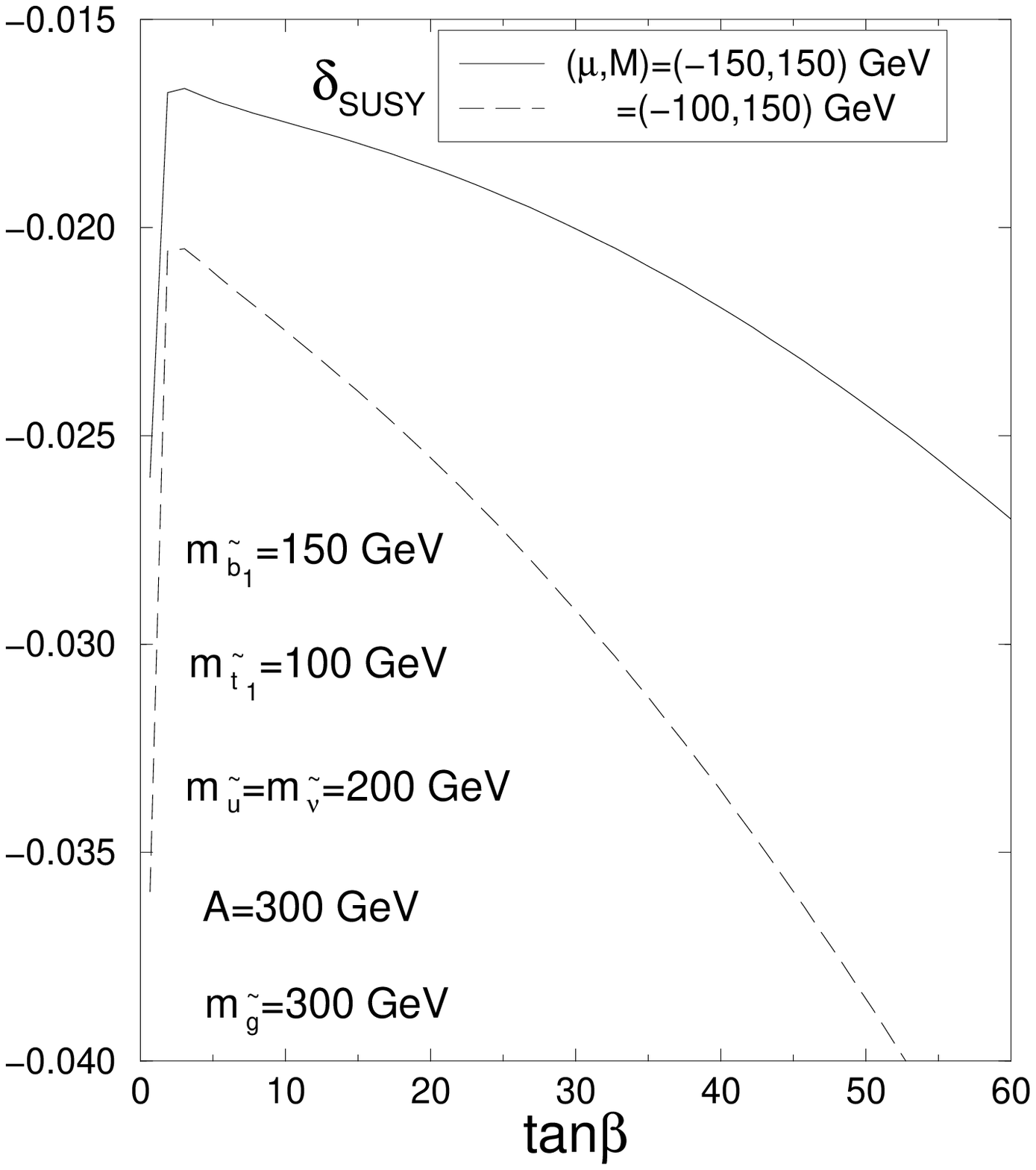,width=5.5cm}}&
          \mbox{\epsfig{file=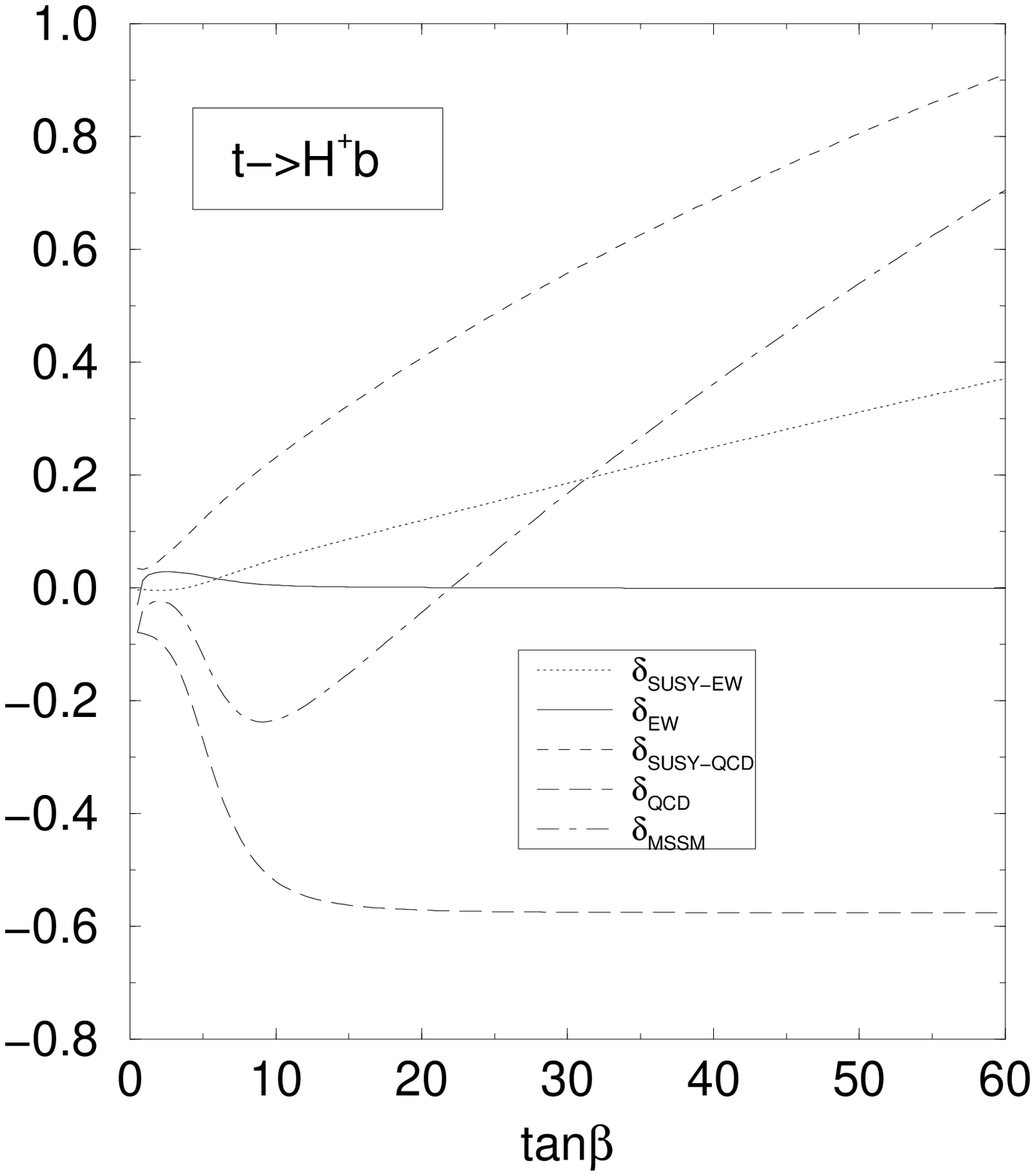,width=5.5cm}} \\(a)&(b)\\[0.5cm]
          \mbox{\epsfig{file=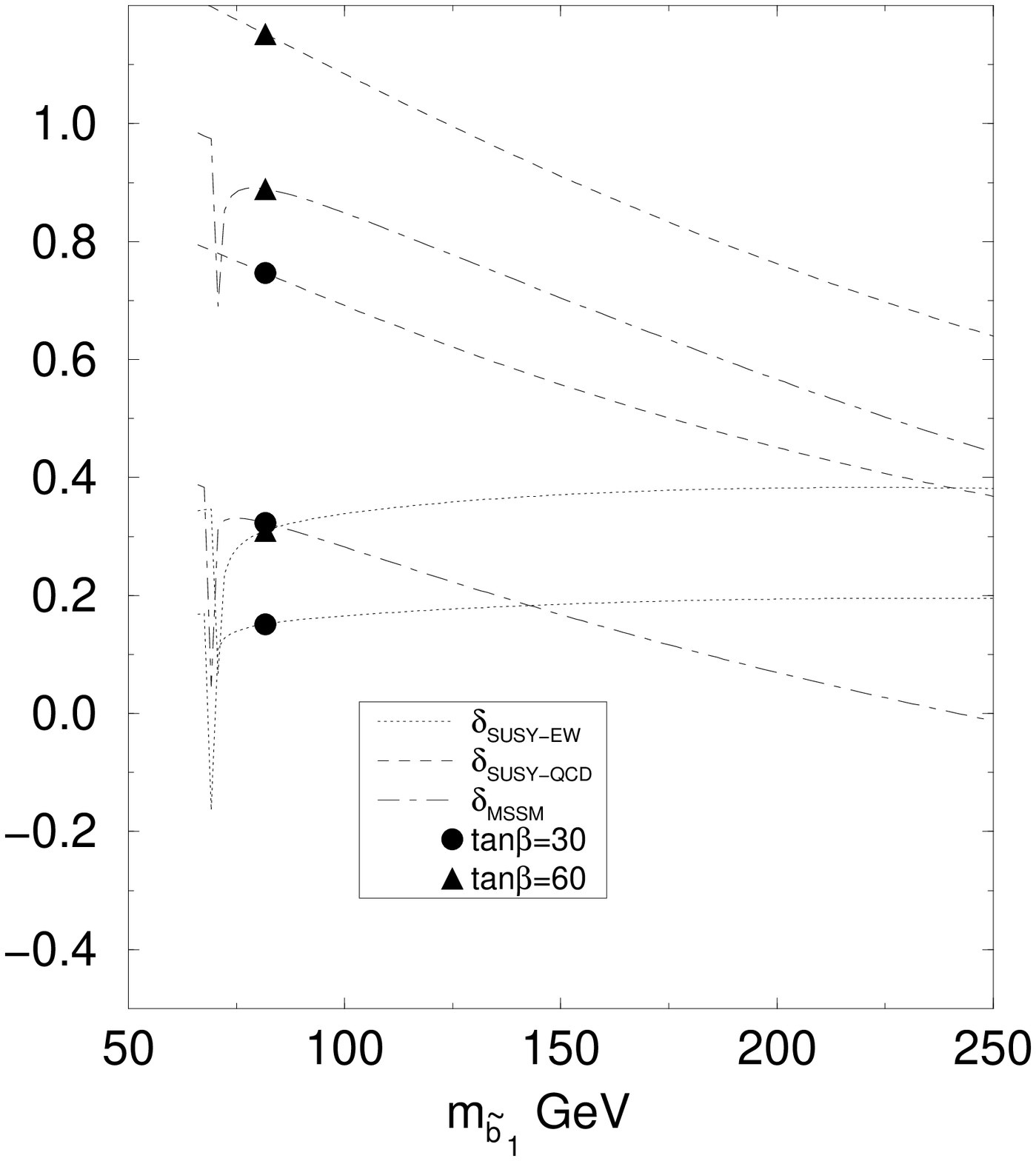,width=5.5cm}}&
          \mbox{\epsfig{file=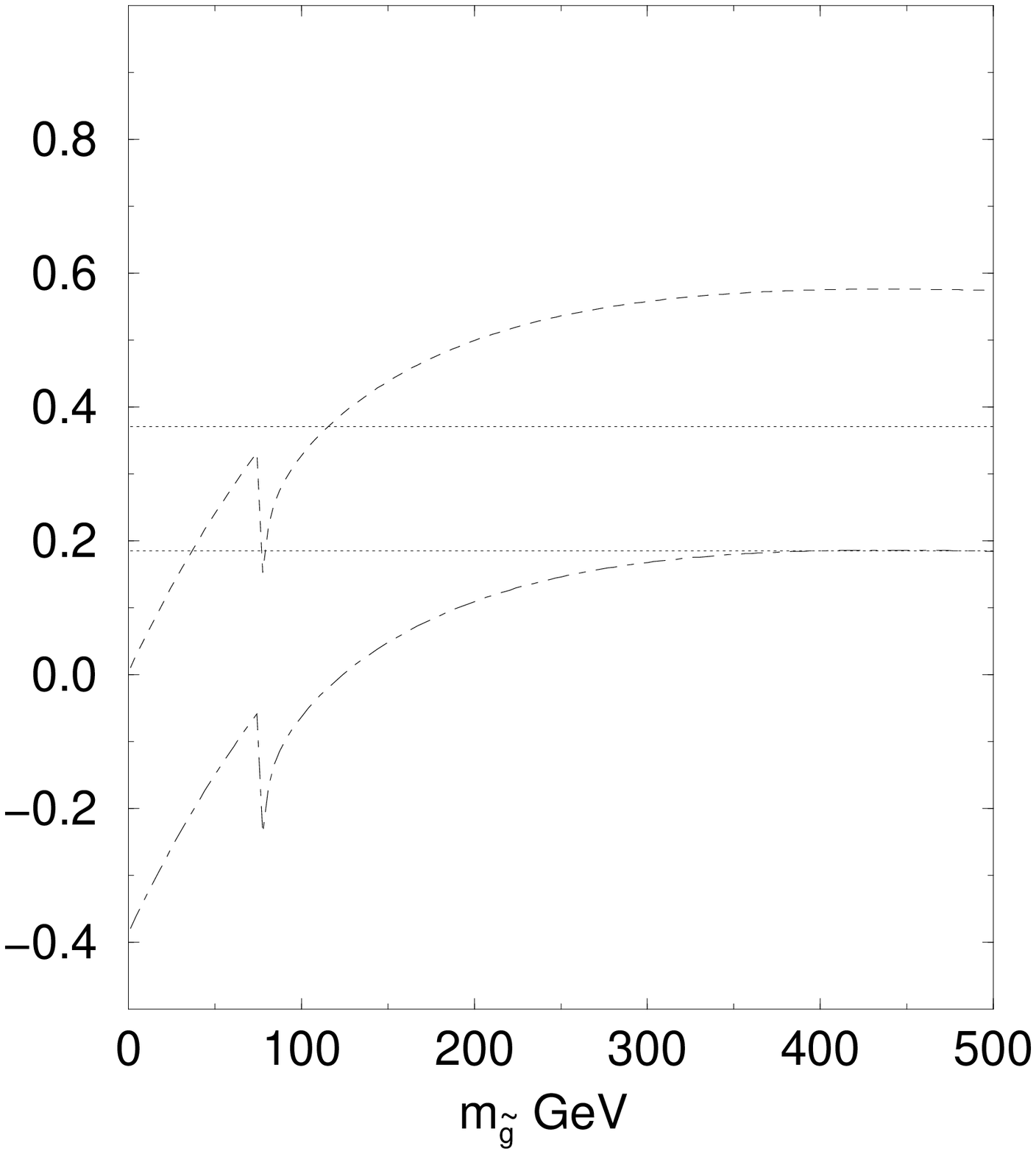,width=5.5cm}} \\(c)&(d)
        \end{tabular}
%%%
%%%%%%  Figures Centrades Verticalment
%%%
%\noindent
\caption
{{\bf (a)} The total SUSY correction to $t\rightarrow W^+\,b$ for given
set of parameters, as a function of $\tan\beta$;
{\bf (b)} The SUSY-EW, standard EW, SUSY-QCD, standard QCD
and full MSSM corrections
as a function of $\tan\beta$. Remaining inputs as in (a);
{\bf (c)} As in (b), but as a function of
$m_{\tilde{b}_1}$; {\bf (d)} As a function of
$m_{\tilde{g}}$.}
%%\end{center}
\end{figure}
%%%%%%%%%%%%%%%%%%%%%%%%%%%%%%%%%%%%%%%%%%%%%%%%%%%%%%%%%%%%%%%%%%%%%%%%%%%% 

The numerical results are conveniently cast in terms of the relative 
correction with respect to the corresponding tree-level
width $\Gamma_0$:
\beq
\delta={\Gamma (\,t\rightarrow (W^+,H^+)\,b\,)
-\Gamma_0 (\,t\rightarrow (W^+,H^+)\,b\,)\over 
\Gamma_0 (\,t\rightarrow (W^+,H^+)\,b\,)}\,.
\label{eq:delta}
\eeq
Let us project out the total SUSY correction (\ref{eq:delta}) for decay i),
i.e. the total MSSM correction after subtracting the SM part.
The latter is defined (in the MSSM context) as the one obtained by
decoupling the sparticle effects and leaving only the lightest
CP-even Higgs contribution ($h^0$) in the limit of infinite CP-odd
Higgs mass ($M_{A^0}\rightarrow\infty$)\,\cite{Hunter}.
In the on-shell $G_F$-scheme, which is characterized by
the set of inputs $(G_F, M_W,M_Z,m_f,M_{SUSY},...)$,
the total SUSY correction $\delta_{\rm SUSY}$
is negative and of the order of a few per cent (except in some
unlikely cases\,\cite{GJSH}). In Fig.\, 2a it is shown the
dependence of the corrections on the crucial parameter $\tan\beta$,
for a typical set of parameters. Here $\mu,\,M$ are the
higgsino and $SU(2)$-gaugino mass parameters, respectively,
and $A$ is the value
of a universal trilinear coupling.
In spite of the enhancement at high $\tan\beta$ the 
negative shift of the 
decay amplitude is below $4\%$ even for $\tan\beta=50$.
Therefore, for $1\stackm\tan\beta\stackm 30$ the negative SUSY effects 
approximately cancel out
against the (positive) electroweak SM contributions, which
are of the same order of magnitude ($\stackm +2\%$), 
leaving the ordinary QCD effects\,\cite{TopSM} 
($\simeq -10\%$) as the net MSSM corrections. As a result no 
significant imprint of the underlying SUSY dynamics is left behind
the standard top quark decay $t\rightarrow W^+\,b$ and we
are thus led to examine other top decays beyond the SM.

In contrast to decay i), decay ii) may receive spectacularly
large SUSY quantum 
corrections, namely of the order of
$50\%$, which certainly could not be missed -- if SUSY is there at all. 
For this reason, we concentrate on that decay. To be sure,
$t\rightarrow H^+\,b$ has been object of many studies 
in the past,
mainly within the context of general 
two-Higgs-doublet models ($2HDM$).
From the experimental point of view that decay 
has been thoroughly scrutinized at the 
Tevatron\,\cite{CDF}. Recently a systematic study
has been made on multilepton and multijet signatures
for the charged Higgs decay of
the top quark at the Tevatron that could be
useful to constraint the $2HDM$ parameter space\,\cite{Dicus}. 
However, it is shown that
the current CDF data\,\cite{CDF2} on $2$ b-jets and $1$
lepton channel do not pose any real restriction on the charged
Higgs decay of the top quark. 
On the other hand no systematic treatment of the
MSSM quantum effects of the decay $t\rightarrow H^+\,b$ 
existed in the literature until 
the works of Refs.\,\cite{CGGJS} and \cite{Guasch4}.
Moreover, remember that in the MSSM, in contrast to the general
$2HDM$, the charged Higgs can elude the stringent
lower mass bounds following from
radiative $B$-decays ($b\rightarrow s\,\gamma$)\,\cite{CLEO}
that would preclude the decay under consideration.
Admittedly, the situation with radiative $B$-decays is
not completely clear since 
there are many sources of error that deserve
further experimental consideration. 
Still this information can be used to single out the SUSY
nature of the Higgs sector.
Thus in the MSSM the existence of decay ii) is 
more tenable than in the framework of an 
unconstrained $2HDM$. 
Next we briefly review the results concerning the 
important MSSM quantum
corrections potentially affecting its decay width.

We present our results for the decay ii) also in the on-shell scheme.
In considering the various parameter dependences,
again a fundamental parameter to be tested 
is $\tan\beta$. 
This parameter is involved explicitly in
the relevant interaction Lagrangian for the decay ii), namely
\beq
{\cal L}_{Htb}={g\over\sqrt{2}M_W}\,H^+\,\bar{t}\,
[m_t\cot\beta\,P_L + m_b\tan\beta\,P_R]\,b+{\rm h.c.}\,,
\label{eq:LtbH}
\eeq 
where $P_{L,R}=1/2(1\mp\gamma_5)$ are the chiral projector operators.
Therefore, crucial for the treatment of the  SUSY-EW effects 
on the decay ii) is the
definition of $\tan\beta$ beyond the tree-level.
Following Ref.\cite{CGGJS} we define it by
means of the $\tau$-lepton decay of $H^\pm$:
\beq
\Gamma(H^{+}\rightarrow\tau^{+}\nu_{\tau})=
{\alpha m_{\tau}^2\,M_H\over 8 M_W^2 s_W^2}\,\tan^2\beta\,. 
\label{eq:tbetainput}
\eeq
This definition generates a counterterm
\beqn
{\delta\tan\beta\over \tan\beta}
&=&\frac{1}{2}\left(
\frac{\delta M_W^2}{M_W^2}-\frac{\delta g^2}{g^2}\right)
-\frac{1}{2}\delta Z_H\nonumber\\
&+&\cot\beta\, \delta Z_{HW}+ 
\Delta_{\tau}\,.
\label{eq:deltabeta}
\eeqn    
Here $\Delta_{\tau}$  
comprises the complete set of
MSSM one-loop effects on the $\tau$-lepton decay of $H^\pm$;
$\delta Z_{H}$ and $\delta Z_{HW}$ stand
respectively for the charged Higgs and mixed
$H-W$ wave-function renormalization factors; and the remaining 
counterterms $\delta g^2$ and $\delta M_W$ are the standard 
ones in the on-shell scheme\,\cite{WdeBoer}.

For a typical choice of parameters,
in Fig.\,2b we plot the various contributions to
(\ref{eq:delta}) from SUSY-QCD, SUSY-EW and the MSSM Higgs sector,
as a function of $\tan\beta$.
The standard QCD correction is also shown\,\cite{CD}.
The full MSSM correction $\delta_{\rm MSSM}$
is defined to be the sum of all these
individual contributions. 
As it turns out that for this decay 
$\delta_{\rm MSSM}>>(\Delta r)_{\rm MSSM}$ it follows that
the difference between the results in the $G_F$-scheme and the
$\alpha$-scheme\,\cite{CGGJS}
is not material in this case; hence the bulk of the
effect is already contained in the $\alpha$-parametrization. This
was certainly {\it not} the case with decay i).
In Fig.2c we display the evolution of
the different corrections with $m_{\tilde{b}_1}$; this is a
critical parameter governing the size of the leading (SUSY-QCD)
corrections. Although $\delta_{\rm SUSY-QCD}$ dies away
relatively fast with increasing $m_{\tilde{b}_1}$,
for large sbottom masses
there remains an undampened SUSY-EW component (essentially controled by
$m_{\tilde{t}_1}$) which can be sizeable enough for stop masses
in the few hundred $GeV$.
The decoupling with the gluino mass is
much slower, and with the remarkable property that before 
entering the decoupling regime it has a long
sustained local maximum around $m_{\tilde{g}}=500\,GeV$ (Cf. Fig.\,2d).
Finally we mention that the corrections also increase with 
$A_t$ and $|\mu|$, and change
sign with $\mu$.  Of course, $\delta_{MSSM}\rightarrow 0$
when all sparticle masses increase simultaneously.

%%%%%%%%%%%%%%%%%%%%%%%%%%%%%%%%%%%%%%%%%%%%%%%%%%%%%%%%%%%%%%%%%%%%%%%
%Fig.3
\begin{figure}
\centering
        \begin{tabular}{cc}
          \mbox{\epsfig{file=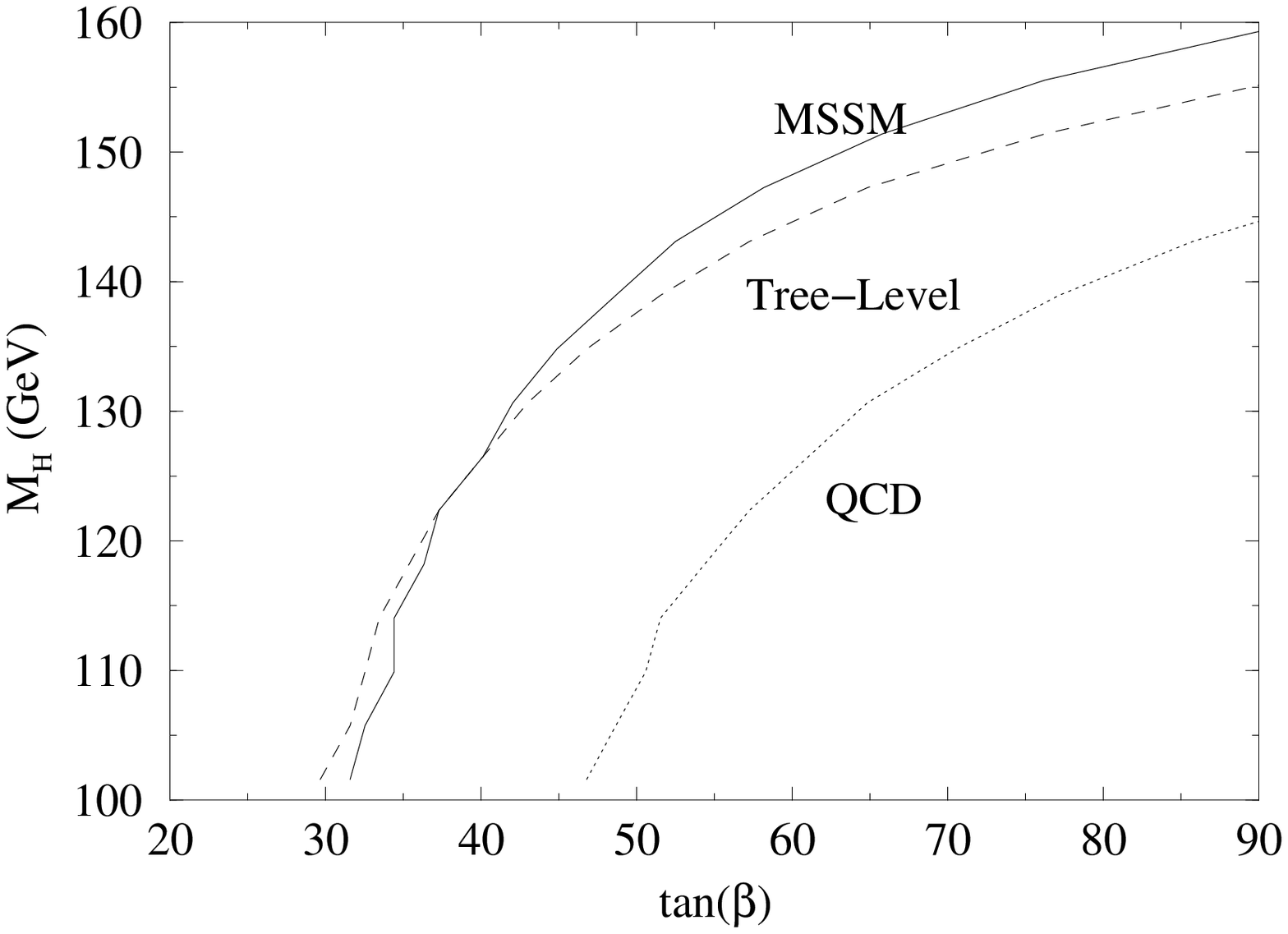,width=5.5cm}}&
          \mbox{\epsfig{file=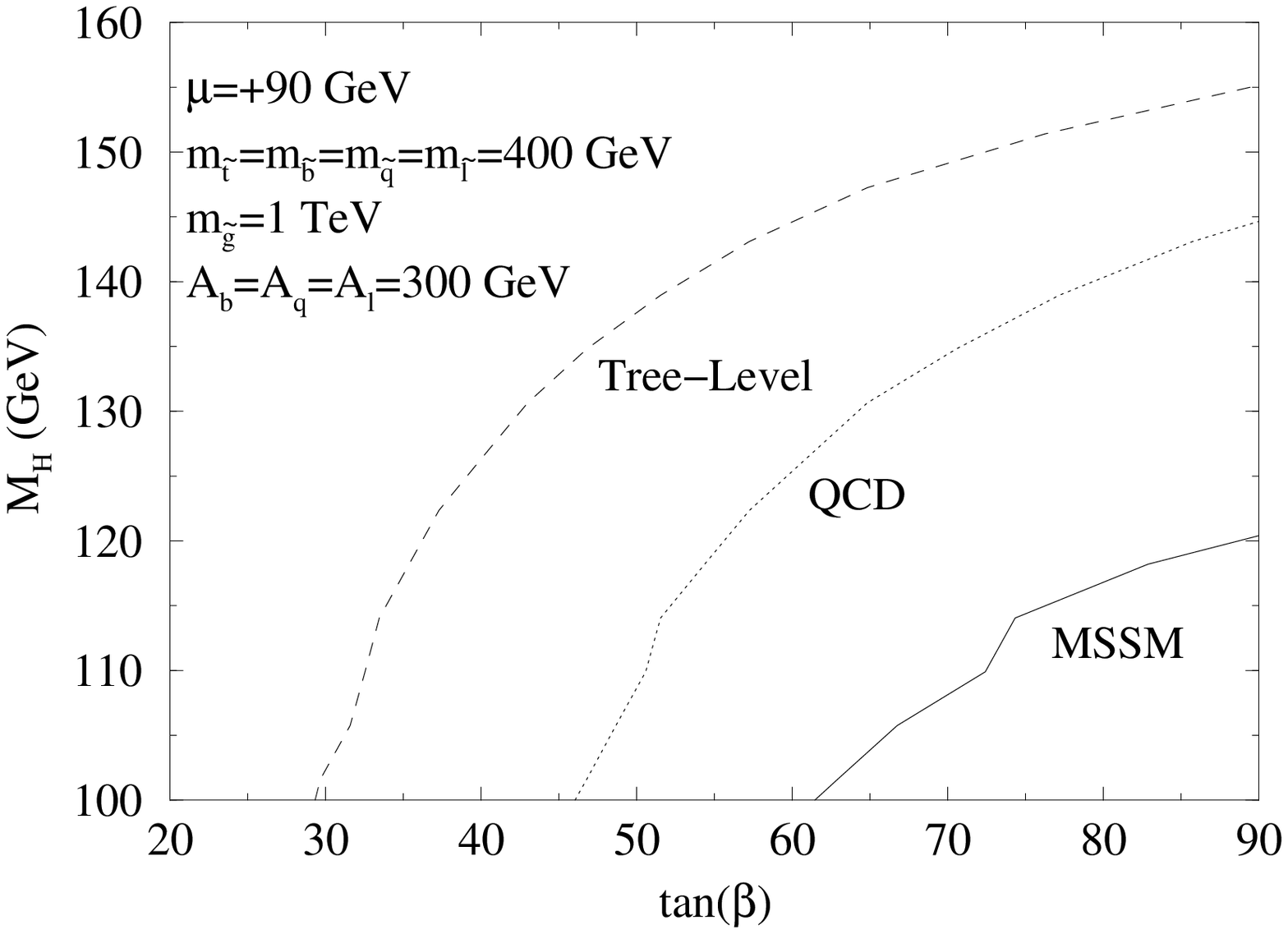,width=5.5cm}} \\(a)&(b)
        \end{tabular}
%%%
%%%%%%  Figures Centrades Verticalment
%%%
%\noindent
\caption
{{\bf (a)} The $95\%$ C.L. exclusion plot in the 
$(\tan\beta, M_{H^\pm})$-plane
for $\mu=-90\,GeV$ and remaining parameters similar to Fig.\,2.
Shown are the tree-level (dashed), QCD-corrected
(dotted) and fully 
MSSM-corrected (continuous) contour lines.
The excluded region in each case is the one lying below the curve;
{\bf (b)} As in (a), but for a $\mu>0$ scenario characterized by a 
heavier SUSY spectrum which makes the analysis compatible
with perturbation theory.}
%%\end{center}
\end{figure}

%%%%%%%%%%%%%%%%%%%%%%%%%%%%%%%%%%%%%%%%%%%%%%%%%%%%%%%%%%%%%%%%%%%%%%%%%%%% 

The definition (\ref{eq:tbetainput}) 
of $\tan\beta$ allows to renormalize the $H^{\pm}\,t\,b$-vertex 
in perhaps the most convenient way to deal with our main decay ii). 
Indeed, from the practical point of view, we should
recall the excellent methods for $\tau$-identification 
developed by the Tevatron
collaborations and recently used by CDF to study the 
existence region of the decay iv) in
the $(\tan\beta,M_H)$-plane\,\cite{CDF}.
However, we wish to show that this analysis may undergo dramatic 
changes when we incorporate the MSSM
quantum effects\,\cite{Guasch4}. Although CDF utilizes
inclusive $\tau$-lepton tagging, for our purposes it will
suffice to focus on the exclusive final state
$(l,\tau)$, with $l$ a light
lepton, as a means for detecting 
an excess of $\tau$-events\,\cite{DPRoy}.
To be precise, we are interested in the
$t\,\bar{t}$ cross-section leading to the decay sequences  
$t\,\bar{t}\rightarrow H^+\,b,W^-\,\bar{b}$ and 
$H^+\rightarrow \tau^+\,\nu_{\tau}$, $W^-\rightarrow l\,\bar{\nu}_l$, 
and {\it vice versa}. 
The relevant quantity can be
easily derived from the measured value
of the canonical cross-section $\sigma_{t\bar{t}}$ for the standard
channel $t\rightarrow b\,l\,\nu_l$, $\bar{t}\rightarrow b\,q\,q'$, after
inserting appropriate branching fractions, namely\,\cite{Guasch4} 
\beq
\sigma_{l\tau}=\left[\frac4{81}\,\epsilon_1+\frac49\,
{\Gamma (t\rightarrow H\,b)\over
\Gamma (t\rightarrow W\,b)}\,\epsilon_2\right]\,\sigma_{t\bar{t}}\,.
\label{eq:bfrac}
\eeq 
The first term in the bracket comes from decay i), and 
for the second term we assume (at high $\tan\beta$) $100\%$ branching 
fraction of $H^+$ into $\tau$-lepton, as explained before. Finally,
$\epsilon_i$ are detector efficiency factors. 
Thus, in most of the phase space available for top decay
the bulk of the cross-section (\ref{eq:bfrac}) is provided by
the contribution of decay ii). Consequently, the 
observable (\ref{eq:bfrac})
should be highly sensitive to MSSM quantum effects.
In fact, from the non-observation of any excess of $\tau$-events
at the Tevatron,
in Figs.\,3a and 3b we derive the ($95\%$ C.L.) excluded regions for
$\mu<0$ and $\mu>0$, respectively. We point out that
the region  of MSSM parameter space considered in this analysis
can be shown to be compatible 
with the $b\rightarrow s\,\gamma$
constraints mentioned above\,\cite{Guasch4}. 
From inspection of these figures
it can hardly be overemphasized that the MSSM quantum effects
can be dramatic. In particular, while for $\mu<0$ the MSSM-corrected
curve is significantly more restrictive than the QCD-corrected one, 
for $\mu>0$ the bound essentially disappears from the perturbative
region ($\tan\beta\stackm 60$). Notice that in the latter case
the SUSY correction is negative and so it adds up to the ordinary
QCD effects. That is why in this 
case we have used a heavier SUSY spectrum than in Fig.\,3a 
in order that the results remain perturbative.

We conclude by pointing out the
recent work of Ref.\cite{Guchait1}. Using Tevatron data in
the $b\bar{b}\tau^+\tau^-$ channel these authors improve the bound
in the $(\tan\beta, M_H)$-space. Nonetheless this analysis was 
performed only at the tree-level and hence it could undergo significant
MSSM radiative corrections. The potentially large
effects not included in that paper
stem from the production mechanism of the CP-odd Higgs boson $A^0$
(through $b\bar{b}$-fusion)
before it decays into $\tau^+\tau^-$ pairs. 
Indeed, the $b\bar{b}\,A^0$ vertex is known\,\cite{Coarasa} to 
develop important MSSM corrections
in the relevant regions of the $(\tan\beta, M_H)$-plane purportedly
``excluded'' by the tree-level analysis of Ref.\cite{Guchait1}.
Clearly, a detailed
re-examination of the excluded region at the quantum level is in order
within the context of the MSSM\,\cite{Guchait2}. 
The lesson to be learnt should be highly instructive: 
namely, in contrast to the tiny corrections to gauge boson 
observables, the MSSM
quantum effects on top-Higgs boson physics can be rather large and
should not be neglected in future searches at the Tevatron and 
at the LHC.

%%%%%%%%%%%%%%%%%%%%%%%%%%%%%%%%%%%%%%%%%%%%%%%%%%%%%%%%%%%%%%%%%%%%%%%%%%%%%%
%\vspace{1cm}
{\bf Acknowledgements}:
The author is thankful to Toni Coarasa, David Garcia, Jaume Guasch
and R.A. Jim\'enez for a fruitful collaboration.  This work
has been partially supported by CICYT under project No. AEN93-0474.
\noindent

%%%%%%%%%%%%%%%%%%%%%%%%%%%%%%%%%%%%%%%%%%%%%%%%%%%%%%%%%%%%%%%%%%%

\end{document}